# Integrated LNOI single-mode lasers by Vernier effect

Ru Zhang,[1] Chen Yang,[1] Zhenzhong Hao,[1] Di Jia,[1] Qiang Luo,[1] Dahuai Zheng,[1] Hongde Liu,[1] Xuanyi Yu,[1] Feng Gao,[1] Fang Bo,[1,2,3,*] Yongfa Kong,[1,4] Guoquan Zhang,[1,2,5] and Jingjun Xu[1,2,6]

[1]MOE Key Laboratory of Weak-Light Nonlinear Photonics, TEDA Institute of Applied Physics and School of Physics, Nankai University, Tianjin 300457, China
[2]Collaborative Innovation Center of Extreme Optics, Shanxi University, Taiyuan 030006, China
[3]Collaborative Innovation Center of Light Manipulations and Applications, Shandong Normal University, Jinan 250358, China
[4]e-mail: kongyf@nankai.edu.cn
[5]e-mail: zhanggq@nankai.edu.cn
[6]e-mail: jjxu@nankai.edu.cn
*Corresponding author: bofang@nankai.edu.cn

**Microcavity lasers based on erbium-doped lithium niobate on insulator (LNOI), which are key devices for LNOI integrated photonics, have attracted much attention recently. In this Letter, we report the realization of a C-band single-mode laser using Vernier effect in two coupled Erbium-doped LNOI microrings with different radii under the pump of a 980-nm continuous laser. The laser, operating stably over a large range of pumping power, has a pump threshold of ~200 µW and a side-mode suppression ratio exceeding 26 dB. The high-performance LNOI single-mode laser will promote the development of lithium niobate integrated photonics.**

In recent years, lithium niobate integrated optical devices activate worldwide attention [1, 2]. On the one hand, it is due to the excellent optical properties of lithium niobate (LN), such as wide transparent window (0.35-5 µm); low intrinsic absorption; extraordinary second-order nonlinear ($d_{33}$=-41.7 pm/V), electro-optic ($r_{33}$=30.9 pm/V), acousto-optic effects; and marked photorefractive, piezoelectric, pyroelectric effects. On the other hand, the presence of lithium niobate on insulator (LNOI) offers an improvement in optical confinement both in time and in space compared with traditional lithium niobate devices fabricated through proton exchange (PE) or indiffusion of titanium. Moreover, the commercialization of LNOI has promoted the development of integrated lithium niobate devices. A large number of passive integrated optical devices including nonlinear frequency converters[3-6], optical parametric oscillators[7], electro-optic/acousto-optic modulators[8-11], and mode converters[12] have been fabricated on LNOI and have achieved excellent performance. Very recently, lasers[13-19] and amplifiers[20-23] based on erbium-doped LNOI have been reported, which greatly promote the research of the active devices based on LN, but their integration and performance need to be further improved.

As a typical microlaser, whispering gallery mode (WGM) microcavity lasers based on microsphere, microbottle, microdisk, micro-toroid, microring, etc. are widly investigated. Compared with the distributed feedback (DFB) laser, quantum well laser, and vertical cavity surface emitting lasers (VCSEL), WGM microcavity lasers via total internal reflection, exhibit higher modal confinement and higher quality ($Q$) factor, thus have lower laser threshold. In virtue of integration, ring cavities integrated with bus waveguides have better performance. Not long ago, our group reported LNOI microring cavity laser[15]. Compared with the disk cavities, the mode confinement in space has a great improvement, and its pump threshold decreased significantly to 20 µW. However, it presents multi-mode operation, which is undesirable for applications such as optical communications and nonlinear optical investigations.

There are several methods to realize single-mode lasers. One is to shape the spatial profile of the optical pump into a laser cavity[24]. Another is to make the high-order laser modes suppressed by pairing it with the losing mode to achieve a single mode[25]. Using parity-time (PT) symmetry to implement modes selection[26-28] is also available as a possibility. Furthermore, increasing the free spectral range (FSR) of the WGM microcavities is a feasible solution. It is well known that the FSR of WGM is inversely proportional to the size of the resonator. To achieve a large FSR, a cavity with a radius of several microns is often required, which inevitably increases the optical loss, deteriorates the laser $Q$ factor, and ultimately leads to an increase in the pump threshold. An effective alternative is to use multiple resonators coupled with each other with different radii to improve the FSR of the system via the Vernier effect[16, 17, 29-31].

In this Letter, we designed and fabricated an LNOI photonic molecule with two coupled ring resonators, whose FSR is about 11 nm. In this device, we obtained a 1531.1-nm single-mode laser with a stable output and a high side-mode suppression ratio (SMSR) of 26 dB under the pump of 980-nm continuous laser. The laser can be flexibly integrated with various functional LNOI devices on a chip.

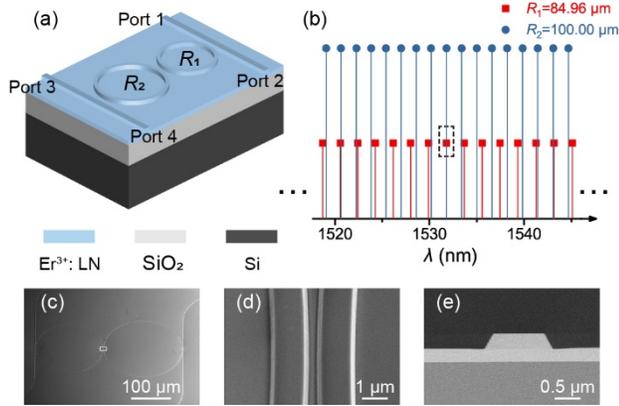

**Fig. 1.** (a) Schematic diagram of the LNOI photonic molecule. $R_{1(2)}$ is the ring radius of the ring resonator 1 (2). The ports of the chip are labeled as port 1, port 2, port 3, and port 4. (b) Numerical simulation results of resonant wavelengths of ring resonators. The black dotted box indicates that the two ring resonators have the same resonance wavelength, that is 1531.8 nm. (c) SEM image of the fabricated LNOI photonic molecule with coupled waveguides. (d) Enlarged view of the coupling region highlighted by the white box in the middle of (c). (e) SEM image of the end face of a bus waveguide.

The basic configuration of the LNOI photonic molecule is shown in Fig. 1(a). It consists of two single-ring resonators with radii of $R_1$ and $R_2$, respectively, and two bus waveguides used to couple the pump light and extract optical signals. The light with wavelengths of $\lambda_i$ resonantes in a single ring resonator when the resonant condition ($2\pi n_i R_i = m_i \lambda_i$, $i=1,2$) is satisfied, in which $m_i$ is an integer, and $n_i$ is the effective refractive index of the mode. The interval between adjacent resonant wavelengths of the mode is $\Delta\lambda_i = \lambda_i^2/(2\pi n_{gi} R_i)$, where $n_{gi}$ is the mode group index. For photonic molecules with two coupled microcavities, the proper choice of $R_1$ and $R_2$ can make the two ring resonators have the same resonant wavelength $\lambda_0 = \lambda_2 = \lambda_0$. In this case, supermodes with $\Delta\lambda \approx \lambda_0^2/(2\pi n_g |R_1-R_2|)$ are supported in the photonic molecule. If $R_1 \approx R_2$, then the $\Delta\lambda$ approaches infinity. Therefore, a relatively large FSR can be achieved in a photonic molecule with relatively large radius supporting high $Q$ factors. To achieve simultaneous resonance in two ring resonators near 1532 nm, at which Erbium ions ($Er^{3+}$) demonstrates strongest radiation, we fixed the radius of one of the large resonator at 100 μm and fine-tuned the radius of the small ring resonator near 85 μm.

We numerically calculated the respective resonant wavelengths of the two resonators by a two-dimensional axisymmetric model. Simulation results showed that when the radius of the small cavity was 84.96 μm, dual resonance could be realized at 1531.8 nm, as shown in Fig. 1(b). Considering the fabrication error, we adjust the radius of the small ring from 84.90 μm to 85.02 μm with a step of 0.02 μm in the fabrication. In this case, the $\Delta\lambda$ was estimated to be ~11 nm.

Erbium-doped X-cut LNOI wafer with a doping concentration of 0.1 mol% was used to fabricate the photonic molecule. The thickness of LN film, silicon-dioxide ($SiO_2$) buffer layer, and silicon (Si) substrate are 0.6 μm, 2 μm, and 500 μm, respectively. The fabrication process of the LNOI photonic molecule is similar to that in our previous paper[15]. The etching depth of the LN film is 300 nm, the ring resonators and the waveguides have a ridged shape with a sidewall angle of ~60°. The shape of the fabricated LNOI photonic molecule was first characterized by using scanning electron microscope (SEM). Figures 1(c-e) show the SEM images of a typical LNOI photonic molecule. As designed, the ring radii of the small ring and the large ring are ~85 μm and 100 μm, respectively. The top width of the rib of the rings is 1.2 μm, and the gap between the two rings is 0.65 μm. The width of the bus waveguide with port 1 and port 2 (port 3 and port 4) is 1 μm (0.6 μm), and the gap between the waveguide and the small (large) ring is 0.65 μm (0.4 μm).

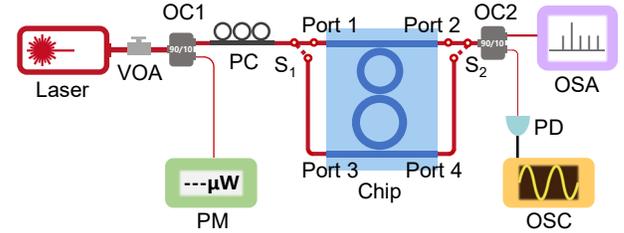

**Fig. 2.** Experimental setup to measure the transmission and lasing performance of the LNOI photonic molecule. VOA: variable optical attenuator; OC: optical coupler; PC: polarization controller; PM: power meter; PD: photodetector; OSA: optical spectrum analyzer; OSC: oscilloscope. The red lines represent optical fibers and the black lines indicate electrical wires.

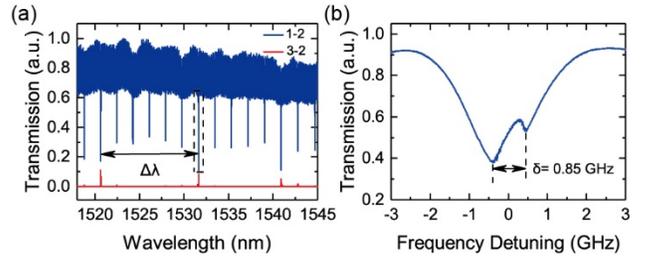

**Fig. 3.** (a) Transmission spectra of an erbium-doped LNOI photonic molecule in 1550-nm band. The blue and red lines indicate the transmitted spectra observed from port 1 to port 2 and from port 3 to port 2, respectively. (b) Transmission spectrum of the mode marked by the dashed lines in (a) shows a mode splitting due to the coupling between the two rings.

To characterize the linear optical properties of the LNOI photonic molecule near 1532 nm, a tunable laser in the band of 1550 nm was used as the light source. The specific experimental setup is shown in Fig. 2. Firstly, we connected the optical fiber at $S_1$ to port 1 on the chip, and connected the optical fiber at $S_2$ to port 2. The laser carried out frequency sweep in the range of 1518-1545 nm, and the transmission spectrum was collected through photodetector (PD) and oscilloscope (OSC). After that, the optical fiber at $S_1$ was switched to port 3, and its transmission spectrum was still observed at port 2 in the same wavelength range.

We tested the designed double-ring structure one by one, and finally, in the photonic molecule structure with $R_1$=85.02 μm, double resonance can be achieved at 1531.6 nm. The transmission spectra from port 1 to port 2 (the blue lines) and from port 3 to port 2 (red lines) are shown in Fig. 3(a). From the transmission spectrum in red, we can see three distinct peaks with ~11-nm $\Delta\lambda$ corresponding to the supermodes of the photonic molecule, which agrees with the theoretical results. The same information can be obtained from the blue transmission spectrum from port 1 to port 2, that is, when the resonant wavelengths of the two cavities are the same, the mode coupling between the waveguide and the photonic molecule is deeper. Figure 3(b) is the enlarged view of the mode highlighted by the dashed black box in Fig. 3(a), from which two supermodes with a splitting $\delta$=0.85 GHz were observed. The frequency interval between the two supermodes is determined by coupling strength between the rings [32].

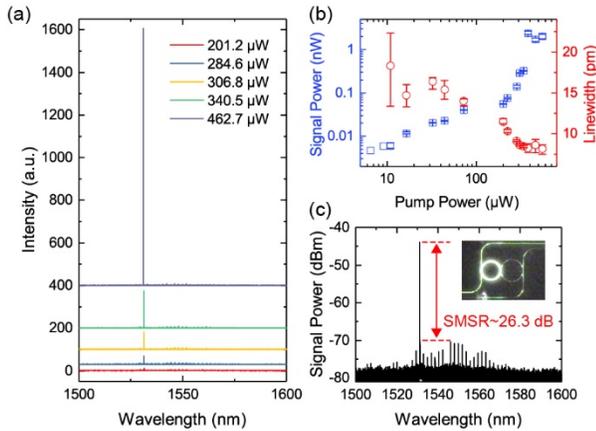

**Fig. 4.** (a) The single-mode lasing in the range of 1500-1600 nm from an erbium-doped LNOI photonic molecule under different pump power. (b) Power (blue hollow box) and linewidth (red hollow circle) of the single-mode lasing signal at different pump power. (c) A high SMSR lasing signal observed at a pump power of ~900 μW. Inset: a micro-images showing the strong green up-conversion fluorescence in the photonic molecule.

We studied the laser emission from the erbium-doped LNOI photonic molecule using the experimental setup shown in Fig. 2 as well. In this case, a 980 nm-band tunable laser was used as the pump. The variable optical attenuator (VOA) was used to change the pump power entering the double-ring resonator. In addition, the optical coupler 1 (OC 1) was required, and 90% of the pump passed through the polarization controller and entered the chip. The remaining part (10%) was sent to the power meter (PM) to monitor the pump power entering the LNOI photonic molecule. The coupling efficiency between the lensed fiber and the waveguide are tuned by three-axis piezo stages. The lensed fiber at $S_1$ is connected to port 1, and the lensed fiber at $S_2$ is connected to port 2 on the chip. Both the signal and the transmitted pump were divided into two parts through the optical coupler 2 (OC 2) after the extracted lensed fiber. Most of the collected light (90%) went into an optical spectrum analyzer (OSA) to obverse the 1550-nm band signal. And a small part of the collected light (10%) was sent to a photodetector (PD), which is connected to an oscilloscope (OSC) to monitor the transmission spectrum of the pump mode around 980 nm.

In experiments, considering the absorption of erbium ions and the maximum tuning wavelength of our laser, the WGM for the pump we selected was at around 979.6 nm. The pump mode has a loaded $Q$ factor of $2.97\times10^5$ and operates in the under-coupling regime. Figure 4(a) shows a series of lasing spectra with various pump power, which clearly show that stable single-mode operation can be achieved in the wavelength range from 1500 nm to 1600 nm over a wide range of pump power. Due to the photorefractive effect of LN, the signal wavelength changes from 1531.4 nm at low power to 1531.1 nm at high power[33].

We performed power integrating and Lorentz fitting for a range of 0.2 nm near the signal peak in the emission spectrum, and obtained the signal power and linewidth. The experimental results are shown in Fig. 4(b). The blue hollow box shows signal power as a function of the pump power, showing two increasing regions, under the log-log coordinate. The first region is the linearly increasing region, and the second region is the superlinearly growing region[34, 35]. In theory, there should be another linear region, and these three regions together form an S-shaped curve. Due to the limitation of the maximum power of the pump laser, we did not collect the S-shaped relationship between the pump power and the signal power. As the pump power increases, the signal linewidth decreases gradually, and eventually fluctuates around a certain value, as shown in the red hollow circle in Fig. 4(b) with log-linear coordinate. Due to the limited resolution of OSA (~10 pm), the linewidth information in the figure at high power is not an accurate representation of the actual linewidth of the laser signal in this case. In fact, it might be much smaller [16, 17]. In Fig. 4 (b), the kinks of the signal power and linewidth with the pump power tend to be the same, accordingly the laser threshold is estimated to be 200 μW. This is an order of magnitude higher than our previous work in microring lasers[15] because of lower $Q$ factors and weak coupling of the pump.

Figure 4(c) shows the single-mode lasing signal in logarithmic coordinates, which was collected at 979.9-nm pump wavelength and 900-μW pump power. The power of the main mode and the highest side mode of the laser are -44.0 dBm and -70.3 dBm, respectively, and the corresponding SMSR is 26.3 dB. The strong green up-conversion fluorescence of the LNOI photonic molecule, in this case, is shown in the inset in Fig. 4(c). At this pump wavelength, only the ring resonator with small radius meets the resonance condition, and the pump energy is mainly concentrated in the small ring resonator, but the signal mode exists in both two ring resonators. In addition to the photonic molecule mode at 1531.6 nm, the photonic molecule mode also exists at 1520.6 nm and 1540.9 nm, as shown in Fig. 3(a). However, due to the low emission coefficient of erbium ion at the other two wavelengths, the laser conversion efficiency is low and the corresponding laser signal is weak, thus we can obtain the laser signal with high SMSR.

In summary, we fabricated the photonic molecule on erbium-doped LNOI and demonstrated a 980-nm-laser-pumped stable C-band single-mode laser by Vernier effect. The lasing threshold is ~200 μW. When the pump power is 900 μW, the SMSR of the single-mode laser reaches ~26.3 dB. In the future, electrodes can be integrated into the photonic molecule, and a tunable single-mode laser can be realized by electro-optic modulation or thermo-optic modulation, which is more convenient for practical application. High-performance single-mode laser may promote the development of the LNOI integrated photonics.

**Funding.** This work was supported by the National Key Research and Development Program of China (Grant No. 2019YFA0705000), the National Natural Science Foundation of China (Grant Nos. 12034010, 11734009, 92050111, 12074199, 92050114, 12004197, and 1774182), and the 111 Project (Grant No. B07013).